\documentclass[twocolumn,prb]{revtex4}
\usepackage{amsfonts}
\usepackage[T1]{fontenc}
\usepackage{amsmath,amsbsy,amssymb,graphicx}
\usepackage{times}
\usepackage{color}
\let\mathbf=\boldsymbol

\begin{document}

\title{Topological quantum quench dynamics carrying arbitrary Hopf and
second-Chern numbers}
\author{Motohiko Ezawa}
\affiliation{Department of Applied Physics, University of Tokyo, Hongo 7-3-1, 113-8656,
Japan}

\begin{abstract}
A quantum quench is a nonequilibrium dynamics governed by the unitary
evolution. We propose a two-band model whose quench dynamics is characterized by an
arbitrary Hopf number belonging to the homotopy group $\pi _{3}(S^{2})=\mathbb{Z}$. 
When we quench a system from an insulator with the Chern number 
$C_{i}\in \pi _{2}(S^{2})=\mathbb{Z}$ to another insulator with the Chern
number $C_{f} $, the preimage of the Hamiltonian vector forms links having
the Hopf number $C_{f}-C_{i}$. We also investigate a quantum-quench dynamics for a
four-band model carrying an arbitrary second-Chern number $N\in \pi
_{4}(S^{4})=\mathbb{Z}$, which can be realized by quenching a
three-dimensional topological insulator having the three-dimensional winding
number $N\in \pi _{3}(S^{3})=\mathbb{Z}$.
\end{abstract}

\maketitle

\section{Introduction:}

Topological physics has been investigated intensively in this decade. It is
characterized by a topological number quantized for distinct phases.
Topological properties are extensively studied in equilibrium, while they are yet
to be explored in nonequilibrium. One successful example is a
Floquet system\cite{Oka09L,Kitagawa01B,Lindner,Dora,EzawaPhoto,Gold}, where
the external field is oscillating. Quantum quench is another method to
create a nonequilibrium state, where some parameters are suddenly changed,
and afterwards the wave function develops under unitary transformation\cite{Caio,Hu,LZhang,Liou,Qiu,Ueda}.

The Hopf number is described by the homotopy class $\pi _{3}(S^{2})$, which
is a linking number in three dimensions. It is naturally realized in a
two-band system in physical system, since it is characterized by $S^{2}$.
Nontrivial Hopf textures are discussed for cold atoms\cite{Kawaguchi, Hall},
light fields\cite{Kedia} and liquid crystal\cite{Ak}. The topological Hopf
insulator is a three-dimensional (3D) topological insulator possessing a
nonzero Hopf number\cite{Moore,Deng,Duan,DengC,Ken,Xu}. The topological Hopf
semimetal has been proposed, whose Fermi surface contains linked loop nodes\cite{WChen,ZYan,PYChang,EzawaHopf,Hasan}. 
Recently, the Hopf number also
appears in the 2D topological insulator after quench\cite{CWang,Tarn,CYang,PYChang,Yu,PYChang2}. 
It is shown that the Hopf number is $1$ when the system turns from a trivial insulator to a topological insulator
with the Chern number $1$. This topological quantum quench has already been
realized in cold atoms by performing quasimomentum-resolved Bloch-state
tomography for the azimuthal phase\cite{Fla,Fla2,Hau}. There are several
studies on the quench from a trivial insulator to a topological insulators,
while there are few studies on the quench from a topological insulator to a
trivial insulator or the quench from a topological insulator to another
topological insulator.

The second-Chern number was originally introduced in the context of the
time-reversal invariant topological insulators in three dimension\cite{TFT},
which is constructed by the dimensional reduction of 4D topological
insulators characterized by the second-Chern number. Since the second-Chern
number is characterized by the homotopy $\pi _{4}(S^{4})$, it requires 4D space.
However, in quantum quench dynamics, since time introduces an additional dimension, the second-Chern
number can be defined in 4D space-time. Indeed, a quantum quench carrying
the second-Chern number was recently proposed\cite{PYChang2}, where the
system is quenched from a trivial insulator to a topological insulator
indexed by the 3D winding number $\pi _{3}(S^{3})$.

In this paper, we propose a model which is characterized by an arbitrary
Hopf number after quench. For this purpose, we first construct a model
carrying an arbitrary Chern number on square lattice. The dynamics of
the density matrix is analytically solved in this system. We show that the
Hopf number is identical to the difference of the Chern numbers between the
initial and final phases. Finally, we propose a quantum quench dynamics
carrying arbitrary second-Chern numbers.

\section{Model}

We consider a two-band tight-binding model defined on square lattice.
The Hamiltonian is given by 
\begin{equation}
H=\left( 
\begin{array}{cc}
F_{1} & F_{2} \\ 
F_{2}^{\ast } & -F_{1}
\end{array}
\right) ,
\end{equation}
where 
\begin{align}
F_{1} &=t_{1}\left( \cos k_{x}+\cos k_{y}\right) +t_{2}\cos k_{x}\cos
k_{y}-m, \\
F_{2} &=\left( \sin k_{x}+i\sin k_{y}\right) ^{N}
\end{align}
in momentum space, with $N$ being an integer. It has Dirac cones at the $\Gamma $ point $(k_{x},k_{y})=(0,0)$, 
the $M$ point $(\pi ,\pi )$, the $X$
point $(\pi ,0)$ and the $Y$ point $(0,\pi )$. The mass is given by the
diagonal element $F_{1}$ at the Dirac point, which reads 
\begin{equation}
M_{\Gamma }=2t_{1}+t_{2}-m
\end{equation}
at the $\Gamma $ point, 
\begin{equation}
M_{M}=-2t_{1}+t_{2}-m
\end{equation}
at the $M$ point, and 
\begin{equation}
M_{X}=M_{Y}=-t_{2}-m
\end{equation}
at the $X$ and $Y$ points.

\begin{figure}[t]
\centerline{\includegraphics[width=0.45\textwidth]{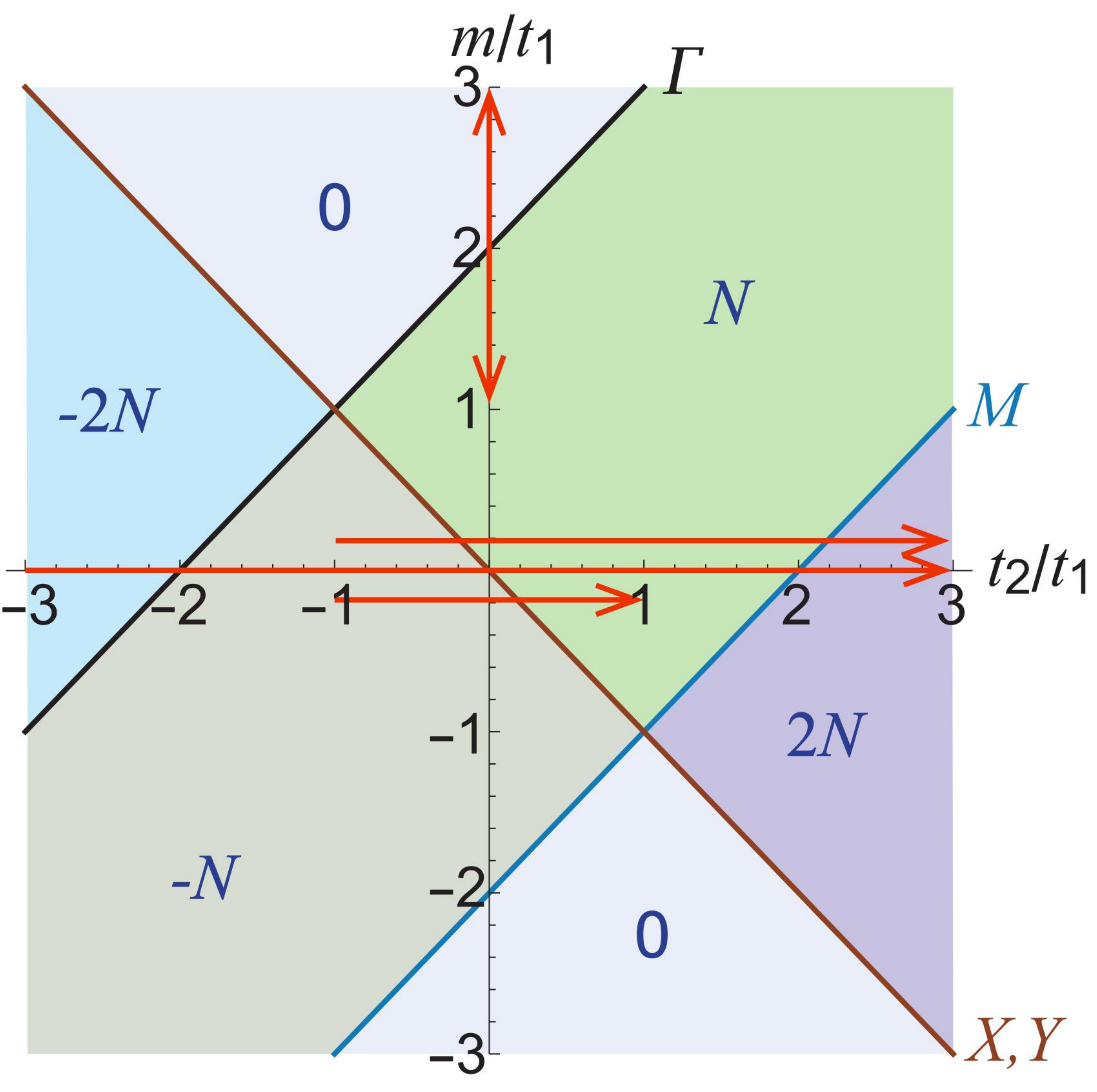}}
\caption{Topological phase diagram as a function of $t_{2}/t_{1}$ and $m/t_{1}$. 
The Chern number is shown in each phase. The phase boundaries are
determined by the condition $M_{K}=0$, where $K$ stands for the
high-symmetry point indicated in the figure. The arrows in red represent the
quantum quench processes we have numerically studied.}
\label{FigPhase}
\end{figure}

There are several topological phases in the Hamiltonian. 
The topological phase diagram is constructed by examining the Dirac masses. 
The phase boundaries are determined by the condions $M_{\Gamma}=M_{M}=M_{X}=M_{Y}=0$ as in Fig.\ref{FigPhase}.

\begin{figure}[t]
\centerline{\includegraphics[width=0.49\textwidth]{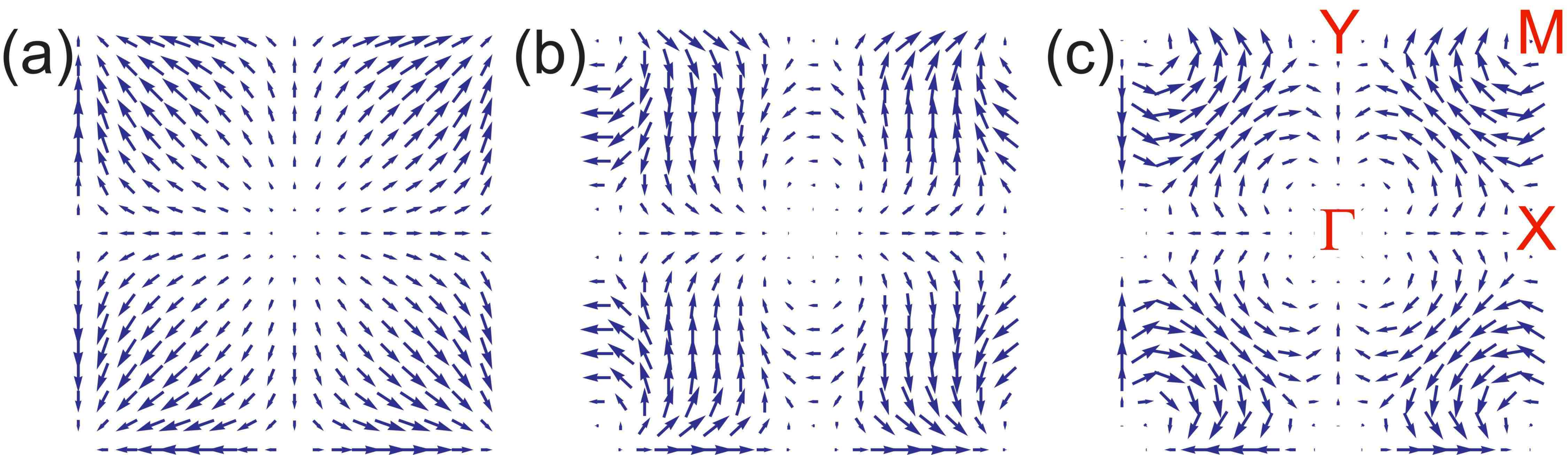}}
\caption{Hamiltonian vector with (a) $N=1$, (b) $N=2$ and (c) $N=3$. They
form meron structure with the winding number $N$ at the high-symmetry points 
$\Gamma ,M,X$ and $Y$,}
\label{FigVector}
\end{figure}

The Hamiltonian is rewritten as 
\begin{equation}
H=\mathbf{d}\cdot \mathbf{\sigma }
\end{equation}
in terms of the Pauli matrices $\mathbf{\sigma }$. 
The normalized vector $\mathbf{\hat{d}=d/}\left\vert \mathbf{d}\right\vert $ points a Bloch sphere
and thus forms an $S^{2}$ manifold. Hence, the Chern number is defined to
characterize the Hamiltonian, which is the Pontryagin number,
\begin{equation}
C=-\frac{1}{4\pi }\int_{\text{BZ}}d^{2}k[\mathbf{\hat{d}}\cdot (\partial _{x}\mathbf{\hat{d}}\times \partial _{y}\mathbf{\hat{d}})].
\end{equation}
We show the $\mathbf{d}$ vector in Fig.\ref{FigVector}. It exhibits meron
structures at the $\Gamma $, $M$, $X$ and $Y$ points. In the vicinity of the
high-symmetry points $K=\Gamma ,M,X$ and $Y$, since $F_{1}=M_{K}$, we may
approximate the Hamiltonian as 
\begin{equation}
H_{K}=\left( 
\begin{array}{cc}
M_{K} & k_{+\xi }^{N} \\ 
k_{-\xi }^{N} & -M_{K}
\end{array}
\right) ,  \label{HK}
\end{equation}
where we have defined $k_{\pm }=k_{x}\pm ik_{y}$ and $\xi =+$ for the $\Gamma $ 
and $M$ points and $\xi =-$ for the $X$ and $Y$ points. 
The $\mathbf{d}$ vector winds $N$ times around the $z$ axis as the azimuthal
angle increases from $0$ to $2\pi $. The total Chern number is thus given by 
\begin{align}
C& =\sum_{K}\xi \frac{N}{2}\text{sgn}\left( M_{K}\right)  \notag \\
& =\frac{N}{2}\text{sgn}\left( 2t_{1}+t_{2}-m\right) +\frac{N}{2}\text{sgn}
\left( -2t_{1}+t_{2}-m\right)  \notag \\
& +N\text{sgn}\left( t_{2}+m\right) .
\end{align}
We show the topological phase diagram in Fig.\ref{FigPhase}. It has five
phases indexed by the Chern numbers $C=0,\pm N,\pm 2N$.

\section{Quantum quench}

We investigate a quantum quench of the Hamiltonian beween two phases in the
phase diagram in Fig.\ref{FigPhase}. We start with an initial Hamiltonian
where $\mathbf{d}=\mathbf{d}^{i}$. At a certain time $t_{0}$, we suddenly
change it to the final Hamiltonian where $\mathbf{d}=\mathbf{d}^{f}$, while
keeping the system to remain in the ground state of the initial Hamiltonian.
(We choose $t_{0}=0$ for simplicity.) After the quantum quench, the system
is no longer the ground state but an excited state with respect to the final
Hamiltonian. For $t>t_{0}$, the dynamics is described by the density matrix 
$\rho _{\pm }=|\psi _{\pm }\rangle \langle \psi _{\pm }|$ satisfying the
Liouville-von-Neumann equation, 
\begin{equation}
i\frac{\partial \rho _{\pm }\left( k,t\right) }{\partial t}=\left[
H^{f}\left( k\right) ,\rho _{\pm }\left( k,t\right) \right] ,
\end{equation}
whose solution is given by an unitary evolution as 
\begin{equation}
\rho _{\pm }\left( k,t\right) =e^{-iH^{f}\left( k\right) t}\rho _{\pm
}\left( k,0\right) e^{iH^{f}\left( k\right) t}.
\end{equation}
In the two-band system, the density matrix is rewritten in terms of the $\mathbf{d}$ vector as 
\begin{equation}
\rho _{\pm }\left( k,0\right) =\left[ 1{\pm }\mathbf{\hat{d}}^{i}\cdot 
\mathbf{\sigma }\right] /2.
\end{equation}
The time-evolved density matrix is then given by\cite{CYang} 
\begin{equation}
\rho _{\pm }\left( k,t\right) =\left[ 1{\pm }\mathbf{\hat{d}}\left(
k,t\right) \cdot \mathbf{\sigma }\right] /2,
\end{equation}
with the time-evolved $\mathbf{d}$ vector 
\begin{equation}
\mathbf{\hat{d}}\left( k,t\right) =\mathbf{e}_{1}+\mathbf{e}_{2}\cos \left(
2\varepsilon t\right) +\mathbf{e}_{3}\sin \left( 2\varepsilon t\right) ,
\label{DVec}
\end{equation}
where we have defined an orthogonal basis\cite{CYang} 
\begin{align}
\mathbf{e}_{1}& =\mathbf{\hat{d}}^{f}\left( \mathbf{\hat{d}}^{i}\cdot 
\mathbf{\hat{d}}^{f}\right) , \\
\mathbf{e}_{2}& =\mathbf{\hat{d}}^{i}-\mathbf{\hat{d}}^{f}\left( \mathbf{\hat{d}}^{i}\cdot \mathbf{\hat{d}}^{f}\right) , \\
\mathbf{e}_{3}& =\mathbf{\hat{d}}^{i}\times \mathbf{\hat{d}}^{f}.
\end{align}
The $\mathbf{d}$ vector is initially $\mathbf{d}^{i}$ and rotates on the
Bloch sphere with period $\pi /\varepsilon $. Hereafter, we rescale the time
as $\tau =\varepsilon t$. Then, the quench dynamics is periodic with period $\pi $ as a function of $\tau $, 
forming a manifold $S^{1}$. The mapping from 
$(k_{x},k_{y},\tau )$ to the Bloch sphere is a mapping from the $T^{3}$ to
the $S^{2}$, which is characterized by the Hopf number.

We study various quantum quench processes from an insulator with $C_{i}$ to
another insulator with $C_{f}$. We first study a quantum quench from a
trivial insulator to a topological insulator, and then a quantum quench from
a topological insulator to a trivial insulator, and finally a quantum quench
from a topological insulator to another topological insulator

\begin{figure}[t]
\centerline{\includegraphics[width=0.45\textwidth]{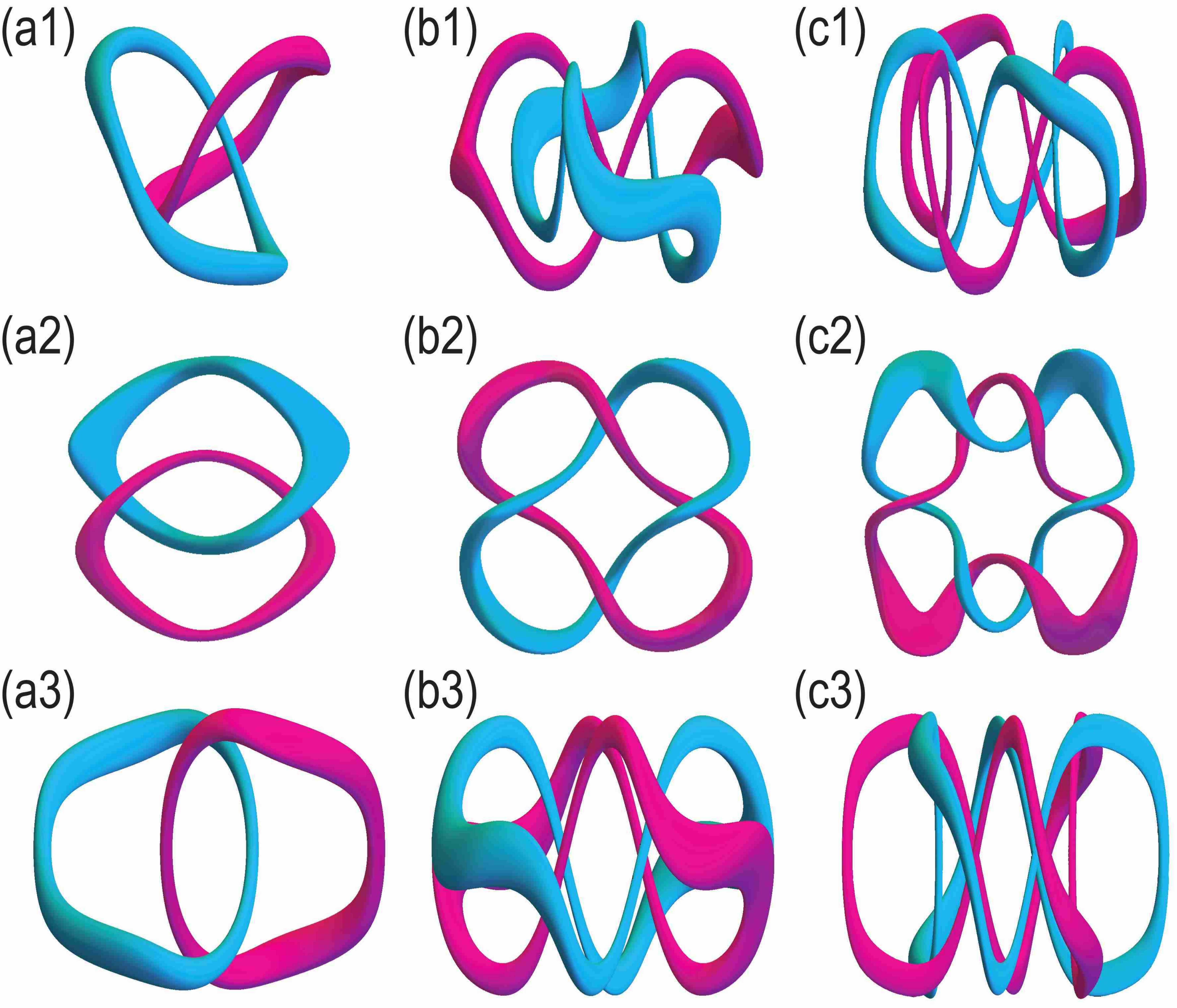}}
\caption{Trivial to topological quench. Bird's eye's view of the almost
zero-energy surface of the Hamiltonian vector with (a1) $N=1$, (b1) $N=2$
and (c1) $N=3$. They form closed linked loops. We have quenched from the
trivial states with the mass $m/t_{1}=3$ to the topological states with the
mass $m/t_{1}=1$ while keeping $t_{2}/t_{1}=0$ to draw figures. The preimage
of $\hat{d}_{y}=1$ is colored in magenta, while that of $\hat{d}_{y}=-1$ is
colored in cyan. (a2)--(c2) and (a3)--(c3) are the corresponding top and
side views. }
\label{FigLink}
\end{figure}

\section{Trivial to topological quench}

First, we consider a quantum quench from a trivial insulator to a
topological insulator with $N$. We show the preimages of $\hat{d}_{y}\left(k_{x},k_{y},\tau \right) =\pm 1$ 
in the space time ($k_{x},k_{y},\tau $) in
Fig.\ref{FigLink}, where the two preimages form torus links with the Hopf number $N$. 
We numerically calculate the quench dynamics between the trivial
insulator with $m/t_{1}=3$ and the topological insulator with $m/t_{1}=1$
while keeping $t_{2}/t_{1}=0$. See the vertical arrow in the phase diagram
(Fig.\ref{FigPhase}), where the vertical arrow crosses only the phase boundary determined by $M_{\Gamma }=0$.

It is possible to construct an analytical expression for a quantum quench from a trivial insulator with the mass $m=\infty $ to a
topological insulator with the mass $M_{\Gamma }$. Note that the trivial
insulators with $m/t_{1}=3$ and $m=\infty $ belong to the same phase. 
In this quench, since the sign of the mass $M_{\Gamma }$ is relevant,
it is enough to analyze the low-energy theory in the vicinity of the $\Gamma $ point, 
where the wave function of the Hamiltonian (\ref{HK}) is given by
\begin{equation}
\left\vert \psi \left( t=0\right) \right\rangle =\frac{1}{c}\left( 
\begin{array}{c}
\left( M_{\Gamma }+\sqrt{k^{2N}+M_{\Gamma }^{2}}\right) k_{+\xi }^{N} \\ 
1
\end{array}
\right)
\end{equation}
with $c$ the normalization constant. Hence, the initial wave function is 
$\left\vert \psi \left( t=0\right) \right\rangle =\left( 1,0\right) ^{t}$ for 
$m=\infty $. By using
\begin{equation}
e^{-iH^{f}\tau }=\left( 
\begin{array}{cc}
\cos \tau -i\hat{d}_{z}\sin \tau & -i\hat{d}_{-}\sin \tau \\ 
-i\hat{d}_{+}\sin \tau & \cos \tau +i\hat{d}_{z}\sin \tau
\end{array}
\right) ,
\end{equation}
the time-evolved wave function is expressed as
\begin{equation}
\left\vert \psi \left( \tau \right) \right\rangle =e^{-iH^{f}\tau
}\left\vert \psi \left( 0\right) \right\rangle =\left( 
\begin{array}{c}
\cos \tau -i\hat{d}_{z}\sin \tau \\ 
-i\hat{d}_{+}\sin \tau
\end{array}
\right) .
\end{equation}
Making a cylindrical symmetric parametrization,
\begin{equation}
\hat{d}_{+}=\frac{k_{+\xi }^{N}}{k^{N}}\sin \theta \left( k\right) ,\qquad 
\hat{d}_{z}=\cos \theta \left( k\right) ,
\end{equation}
with
\begin{equation}
\cos \theta \left( k\right) =\frac{M_{\Gamma }}{\sqrt{k^{2N}+M_{\Gamma }^{2}}},
\quad \sin \theta \left( k\right) =\frac{k^{N}}{\sqrt{k^{2N}+M_{\Gamma}^{2}}},
\end{equation}
we obtain the Berry connection 
\begin{align}
A_{x}& =\sin \tau \sin \theta \left( k\right) [-\frac{Nk_{y}}{k^{2}}\sin
\tau \sin \theta \left( k\right) +\frac{k_{x}}{k}\theta ^{\prime }\left(
k\right) \cos \tau ], \\
A_{y}& =\sin \tau \sin \theta \left( k\right) [\frac{Nk_{x}}{k^{2}}\sin \tau
\sin \theta \left( k\right) +\frac{k_{y}}{k}\theta ^{\prime }\left( k\right)
\cos \tau ], \\
A_{t}& =-\cos \theta \left( k\right) ,
\end{align}
and the Berry curvature 
\begin{align}
F_{x}& =2\sin \tau \sin \theta \left( k\right) [-\frac{Nk_{x}}{k}\cos \tau
\sin \theta \left( k\right) +\frac{k_{y}}{k}\partial _{k}\theta \left(
k\right) \sin \tau ], \\
F_{y}& =-2\sin \tau \sin \theta \left( k\right) [\frac{Nk_{y}}{k}\cos \tau
\sin \theta \left( k\right) +\frac{k_{x}}{k}\partial _{k}\theta \left(
k\right) \sin \tau ], \\
F_{t}& =\frac{N}{k}\partial _{k}\theta \left( k\right) \sin ^{2}\tau \sin
2\theta \left( k\right) .
\end{align}
By using
\begin{equation}
\mathbf{A}\cdot \mathbf{F}=-\frac{2N}{k}\left[ \partial _{k}\cos \theta
\left( k\right) \right] \sin ^{2}\tau ,
\end{equation}
the Hopf number is calculated as
\begin{align}
\chi & =\frac{1}{\pi }\int_{0}^{\pi }d\tau \int_{0}^{\infty }kdk\mathbf{A}
\cdot \mathbf{F}  \notag \\
& =-N\left[ \lim_{k\rightarrow \infty }\cos \theta \left( k\right) -\cos
\theta \left( 0\right) \right]  \notag \\
& =-N\left[ \lim_{k\rightarrow \infty }\frac{M_{\Gamma }}{\sqrt{k^{2N}+M_{\Gamma }^{2}}}-\text{sgn}M_{\Gamma }\right]  \notag \\
& =N\text{sgn}M_{\Gamma }.  \label{H1}
\end{align}
It is identical to the change of the Chern number at the $\Gamma $ point
since the Chern number at the $\Gamma $ point is given by $\frac{N}{2}$sgn$M_{\Gamma }$. 
Namely, the Hopf number is identical to the Chern number $C_{f} $. 

\begin{figure}[t]
\centerline{\includegraphics[width=0.45\textwidth]{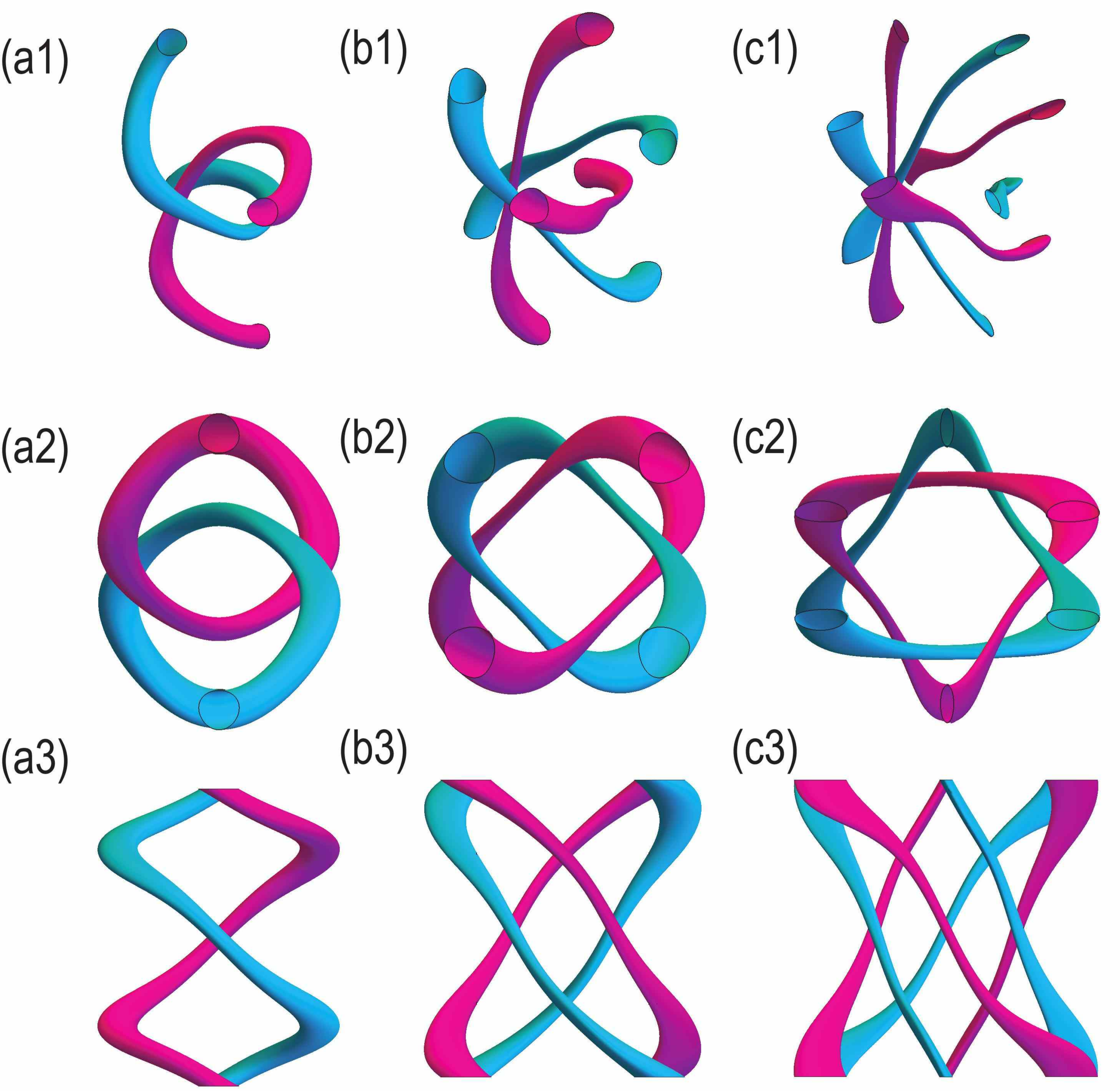}}
\caption{ Topological to trivial quench. Bird's eye's view of the almost
zero-energy surface of the Hamiltonian vector with (a1) $N=1$, (b1) $N=2$
and (c1) $N=3$. They form open linked helix. We have quenched from the
topological states with the mass $m/t_{1}=1$ to the trivial states with the
mass $m/t_{1}=3$ while keeping $t_{2}/t_{1}=0$ to draw figures. The preimage
of $\hat{d}_{y}=1$ is colored in magenta, while that of $\hat{d}_{y}=-1$ is
colored in cyan. (a2)--(c2) and (a3)--(c3) are the corresponding top and
side views. }
\label{FigHelix}
\end{figure}

\section{Topological to trivial quench}

We next consider a quantum quench from a topological insulator with $N$ to
a trivial insulator.
We show the preimages of $\hat{d}_{y}\left(k_{x},k_{y},\tau \right) =\pm 1$ in Fig.\ref{FigHelix}, 
where the two preimages form open helix links with the Hopf number $N$. 
We numerically calculate the quench dynamics between the topological insulator
with $m/t_{1}=1$ and the trivial insulator with $m/t_{1}=3$ while keeping $t_{2}/t_{1}=0$: 
See the vertical arrow in the phase diagram (Fig.\ref{FigPhase}).

It is possible to analytically discuss the Hopf number in an extreme case of
the quantum quench from a topological insulator with the mass $M_{\Gamma }$
to a trivial insulator with the infinite mass $m=\infty $. By inserting the
final state $\mathbf{\hat{d}}^{f}\mathbf{=}\left( 0,0,-1\right) $ into (\ref{DVec}) and we find
\begin{equation}
\mathbf{e}_{1}=\left( 0,0,d_{z}^{i}\right) ,\quad \mathbf{e}_{2}=\left(
d_{x}^{i},d_{y}^{i},0\right) ,\quad \mathbf{e}_{3}=\left(
d_{y}^{i},-d_{x}^{i},0\right) ,
\end{equation}
and
\begin{align}
\mathbf{\hat{d}}\left( k,\tau \right) & =(\cos \left( 2\varepsilon t\right)
d_{x}^{i}+\sin \left( 2\varepsilon t\right) d_{y}^{i},  \notag \\
& \qquad \cos \left( 2\varepsilon t\right) d_{y}^{i}-\sin \left(
2\varepsilon t\right) d_{x}^{i},d_{z}^{i}).
\end{align}
The time-evolved Hamiltonian is proportional to
\begin{equation}
\mathbf{\hat{d}}\left( k,\tau \right) \cdot \mathbf{\sigma =}\left( 
\begin{array}{cc}
\hat{d}_{z} & \hat{d}_{-}e^{2i\tau } \\ 
\hat{d}_{+}e^{-2i\tau } & -\hat{d}_{z}
\end{array}
\right) .
\end{equation}
In the vicinity of the $\Gamma $ point, the wave function is given by
\begin{equation}
\left\vert \psi \left( \tau \right) \right\rangle =\frac{1}{c}\left( 
\begin{array}{c}
-e^{2i\tau }\left( -M_{\Gamma }+\sqrt{k^{2N}+M_{\Gamma }^{2}}\right) \\ 
k_{+}^{N}
\end{array}
\right) .
\end{equation}
The Berry connection is given by
\begin{align}
A_{x}& =-\frac{Nk_{y}}{2k^{2}}\left( 2+\frac{M_{\Gamma }}{\sqrt{k^{2N}+M_{\Gamma }^{2}}}\right) , \\
A_{y}& =\frac{Nk_{x}}{2k^{2}}\left( 2+\frac{M_{\Gamma }}{\sqrt{k^{2N}+M_{\Gamma }^{2}}}\right) , \\
A_{t}& =1-\frac{M_{\Gamma }}{\sqrt{k^{2N}+M_{\Gamma }^{2}}},
\end{align}
and the Berry curvature is given by
\begin{align}
F_{x}& =\frac{NM_{\Gamma }k_{y}}{k^{2N-2}\left( k^{2N}+M_{\Gamma
}^{2}\right) ^{3/2}}, \\
F_{y}& =-\frac{NM_{\Gamma }k_{x}}{k^{2N-2}\left( k^{2N}+M_{\Gamma
}^{2}\right) ^{3/2}}, \\
F_{t}& =\frac{N^{2}M_{\Gamma }}{k^{2N-2}\left( k^{2N}+M_{\Gamma }^{2}\right)
^{3/2}}.
\end{align}
By using
\begin{equation}
\mathbf{A}\cdot \mathbf{F}=-\frac{N^{2}M_{\Gamma }}{k^{2N-2}\left(
k^{2N}+M_{\Gamma }^{2}\right) ^{3/2}},
\end{equation}
the Hopf number is calculated as
\begin{align}
\chi & =\frac{1}{\pi }\int_{0}^{\pi }d\tau \int_{0}^{\infty }kdk\mathbf{A}
\cdot \mathbf{F}  \notag \\
& =-\int_{0}^{\infty }\frac{N^{2}M_{\Gamma }}{k^{2N-2}\left(
k^{2N}+M_{\Gamma }^{2}\right) ^{3/2}}kdk  \notag \\
& =-N\text{sgn}M_{\Gamma }.  \label{H2}
\end{align}
As a result, we find the Hopf number to be $N$. We note that the signs of 
(\ref{H1}) and (\ref{H2}) are opposite since the two processes are the
inverse processes. As in the case of the trivial to topological quench, it
is identical to the change of the Chern number at the $\Gamma $ point since
the Chern number at the $\Gamma $ point is given by $\frac{N}{2}$sgn$M_{\Gamma }$. 
Namely, the Hopf number is identical to the Chern number $-C_{i}$.

\begin{figure}[t]
\centerline{\includegraphics[width=0.49\textwidth]{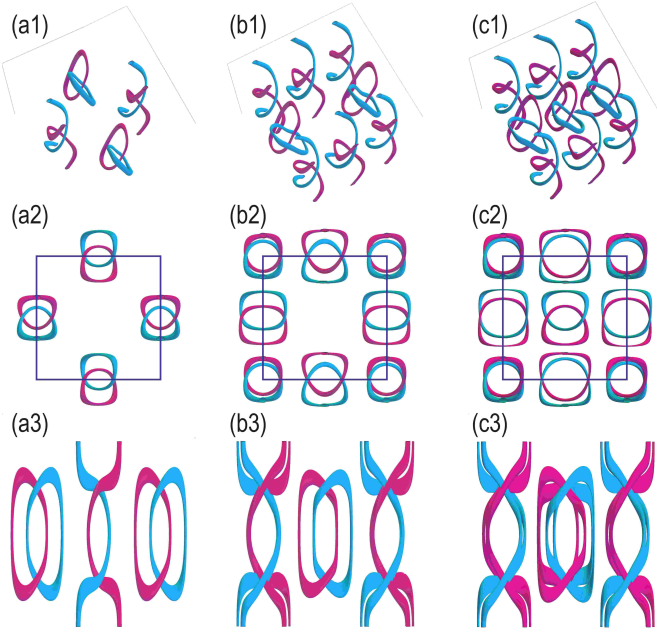}}
\caption{ Topological to topological quench (a) from a topological insulator
with $C_i=-1$ to another topological insulator with $C_f=1$, (b) from a
topological insulator with $C_i=-1$ to another topological insulator with $C_f=2$ 
and (c) from a topological insulator with $C_i=-2$ to another
topological insulator with $C_f=2$. The Brillouin zone is indicated by the
square. The preimage of $\hat{d}_{y}=1$ is colored in magenta, while that of 
$\hat{d}_{y}=-1$ is colored in cyan.}
\label{FigMulti}
\end{figure}

\section{Topological to topological quench}

Finally, we study quench from a topological insulator to another topological
insulator. We show preimages in Fig.\ref{FigMulti} for the case of $N=1$. We
find that torus links and open helix links with the Hopf number $N$ appear at
the high-symmetry points where the sign of the mass changes and the Hopf
number is identical to $C_{f}-C_{i}$. We explicitly study the following
three cases, where the parameter $t_{2}$ is quenched while keeping $m=0$:
See the horizontal arrows in the phase diagram given in Fig.\ref{FigPhase}.
In the following, we study the case with $N=1$ for simplicity, where links
with the Hopf number $1$ emerge at the high-symmetry points. 

(a) For example, if we quench from the topological insulator with 
$t_{2}/t_{1}=-1$ and $C_{i}=-1$ to the topological insulator with 
$t_{2}/t_{1}=1$ and $C_{f}=1$, two Hopf links with the Hopf number $1$ appear
at the $X$ and $Y$ points since the sign of the masses at the $X$ and $Y$
points change [Fig.\ref{FigMulti}(a)]. Then the total Hopf number is $2$,
which is identical to the difference of the Chern numbers. The shape of the
Hopf link at the $X$ point is the closed loop, while that at the $Y$ point
is the open helix. It seems that the $C_{4}$ symmetry is violated. However,
this is an artifact due to the preimages of $\hat{d}_{y}=\pm 1$. The closed
loop and the open helix are inverted when we plot the preimages of $\hat{d}_{x}=\pm 1$.

(b) When we quench from the topological insulator with $t_{2}/t_{1}=-1$ and 
$C_{i}=-1$ to the topological insulator with $t_{2}/t_{1}=3$ with $C_{f}=2$,
three Hopf links with the Hopf number $1$ appear at the $X$, $Y$ and $M$
points [Fig.\ref{FigMulti}(b)]. Then the total Hopf number is $3$.

(c) In the same way, when we quench from the topological insulator with 
$t_{2}/t_{1}=-3$ and $C_{i}=-2$ to the topological insulator with 
$t_{2}/t_{1}=3$ with $C_{f}=2$, three Hopf links with the Hopf number $1$
appear at the $\Gamma $, $X$, $Y$ and $M$ points [Fig.\ref{FigMulti}(c)].
Then the total Hopf number is $4$.

\section{Quantum quench with second-Chern number}

A quantum quench carrying the second-Chern number $1$ has been proposed\cite{PYChang2}. 
We generalize it to a quantum quench carrying an arbitrary
second-Chern number $N$.

We consider the Hamiltonian 
\begin{equation}
H=\sum_{\alpha =x,y,z}f_{\alpha }(\mathbf{k})\tau _{x}\sigma _{\alpha }+m(\mathbf{k})\tau _{z}\sigma _{0},
\end{equation}
where $\tau $ and $\sigma $ represent the Pauli matrices, while $\sigma _{0}$ is the unit matrix. 
We define a unit
vector 
\begin{equation}
d_{\mathbf{k}}=\frac{1}{|E(\mathbf{k})|}(f_{x}(\mathbf{k}),f_{y}(\mathbf{k}),f_{z}(\mathbf{k}),m(\mathbf{k}))
\end{equation}
with the energy 
\begin{equation}
E(\mathbf{k})=\pm \sqrt{\sum_{\alpha =x,y,z}f_{\alpha }^{2}(\mathbf{k})+m^{2}(\mathbf{k})}.
\end{equation}
Since the unit vector forms a three sphere $S^{3}$, the Hamiltonian is
characterized by the 3D winding number\cite{Sch,Volovik} $\nu
_{3}$ describing the third Homotopy $\pi _{3}(S^{3})=\mathbb{Z}$, 
\begin{equation}
\nu _{3}=\frac{1}{2\pi ^{2}}\int_{\text{BZ}}d^{3}k\varepsilon ^{abcd}\hat{d}_{a}
\partial _{k_{x}}\hat{d}_{b}\partial _{k_{y}}\hat{d}_{c}\partial _{k_{z}}
\hat{d}_{d}.  \label{NuThree}
\end{equation}
The unitary evolution is given by $U(t)=\exp [-i\tau H]$. When we start with
the initial state $|\psi \left( 0\right) \rangle =(1,0,0,0)^{t}$, the
quenched wave function is given by 
\begin{align}
\left\vert \psi \left( \tau \right) \right\rangle & =e^{-iH^{f}\tau
}\left\vert \psi \left( 0\right) \right\rangle  \notag \\
& =(\cos \tau ,-im_{\mathbf{k}}\sin \tau ,0,  \notag \\
& \qquad -if_{z,{\mathbf{k}}}\sin \tau ,(f_{y,{\mathbf{k}}}-if_{x,{\mathbf{k}}})\sin \tau )^{t}.
\end{align}
We define the order parameter as 
\begin{equation}
\mathbf{L}=\langle \psi _{\mathbf{k}}(t)|\left( \tau _{x}\sigma _{x},\tau
_{x}\sigma _{y},\tau _{x}\sigma _{z},\tau _{z},\tau _{y}\right) |\psi _{\mathbf{k}}(t)\rangle ,
\end{equation}
which forms the four sphere $S^{4}$ since $\mathbf{L}$ is a unit vector
satisfying $|\mathbf{L}|=1$. It is classified by the fourth homotopy $\pi
_{4}(S^{4})$, where the dynamical second-Chern number is defined by\cite{TFT,PYChang2} 
\begin{equation}
C_{2}=-\frac{3}{8\pi ^{2}}\int_{0}^{\pi /2}\!\!\!\!\!dt\int_{\text{BZ}
}\!\!d^{3}k\varepsilon ^{abcde}L_{a}\partial _{k_{x}}L_{b}\partial
_{k_{y}}L_{c}\partial _{k_{z}}L_{d}\partial _{t}L_{e}.
\end{equation}
It is shown that the dynamical second-Chern number is identical to the 3D
winding number\cite{PYChang2} 
\begin{equation}
C_{2}=\nu _{3}\frac{3}{4\pi ^{2}}\int_{0}^{\pi /2}\sin ^{3}2\tau d\tau =\nu
_{3}
\end{equation}
with (\ref{NuThree}).

Now we explicitly study the model given by 
\begin{align}
f_{x}(\mathbf{k})& =\text{Re}[(\sin k_{x}+i\sin k_{y})^{N}], \\
f_{y}(\mathbf{k})& =\text{Im}[(\sin k_{x}+i\sin k_{y})^{N}], \\
f_{z}(\mathbf{k})& =\sin k_{z}, \\
m(\mathbf{k})& =m-t_{1}(\cos k_{x}+\cos k_{y}+\cos k_{z}).
\end{align}
It follows that the 3D winding number is given by $\nu _{3}=N$ for 
$1<|m/t_{1}|<3$, $\nu _{3}=-2N$ for $|m/t_{1}|<1$ and $\nu _{3}=0$ for $|m/t_{1}|>3$. 
Accordingly, the quantum quench is characterized by the
second-Chern number $N$.

\section{Conclusion}

We have constructed models of quantum quench, which are characterized by an
arbitrary Hopf number or by an arbitrary second-Chern number. We have
explored new types of topological quantum quenches. One is the topological
to trivial quench and the other is the topological to topological quench,
which have different link structures compared to the previously studied
trivial to topological quench.

The author is very much grateful to N. Nagaosa for many helpful discussions
on the subject. This work is supported by the Grants-in-Aid for Scientific
Research from MEXT KAKENHI (Grant No. JP18H03676, No.JP17K05490, and
No.15H05854). This work is also supported by CREST, JST (JPMJCR16F1).

\end{document}